\documentclass[useAMS,usenatbib]{mn2e}
\usepackage{epsfig,graphicx,aas_macros,subfig}
\usepackage{graphics}
\usepackage{amsmath}

\begin{document}
\title{{A \it Suzaku} view of IGR J16393-4643}
\date{}
\author[N. Islam, C. Maitra, P. Pradhan and B. Paul]
{Nazma Islam$^{1,2}$\thanks{E-mail:nazma@rri.res.in},
Chandreyee Maitra$^{3}$, Pragati Pradhan$^{4,5}$ and Biswajit Paul$^{1}$ \\
$^{1}$Raman Research Institute, Sadashivnagar, Bangalore-560080, India\\
$^{2}$Joint Astronomy Programme, Indian Institute of Science, Bangalore-560012, India \\
$^{3}$Laboratoire AIM, CEA-IRFU/CNRS/Universit{\'e} Paris Diderot, Service d'Astrophysique, CEA Saclay, F-91191 Gif sur Yvette, France \\
$^{4}$St. Joseph's College, Singamari, Darjeeling-734104, West Bengal, India \\
$^{5}$North Bengal University, Raja Rammohanpur, District Darjeeling-734013, West Bengal, India}

\maketitle

\begin{abstract}
The pulsar IGR J16393-4643 belongs to a class of highly absorbed supergiant HMXBs, characterised by very high column density of 
absorbing matter. We present the results of the simultaneous broad-band pulsation and spectrum analysis, from a 44 kilosec {\it Suzaku} observation of the source. 
The orbital intensity profile created with the {\it SWIFT-BAT} light-curve shows an indication of IGR J16393-4643 being an eclipsing system with a 
short eclipse semi-angle $\theta_{E} \sim$ 17$^{\circ}$. For a supergiant companion star with a 20 R$_{\odot}$ radius, this implies an inclination of the orbital plane 
in the range of 39$^{\circ}$-57$^{\circ}$, whereas for a main sequence B star as the companion with a 10 R$_{\odot}$ radius, the inclination of the orbital plane is in the 
range of 60$^{\circ}$-77$^{\circ}$. Pulse profiles created for different energy bands, have complex morphology which shows some energy dependence and increases 
in pulse fraction with energy. We have also investigated broad-band spectral characteristics, for phase averaged and resolving the pulse phase into peak phase 
and trough phase. The phase averaged spectrum has a very high N$_{H} (\sim 3 \times 10^{23}$ cm$^{-2}$) and is described by 
power-law ($\Gamma \sim$ 0.9) with a high energy cut-off above 20 keV. We find a change in the spectral index in the peak phase and trough phase, 
implying an underlying change in the source spectrum. 
\end{abstract}


\section{Introduction}

Supergiant High Mass X-ray Binaries (HMXBs) account for one third of the galactic HMXB population. The compact object has an early type supergiant 
star as a companion and the accretion unto the compact object occurs via stellar wind or Roche lobe overflow.
They are subdivided into Roche lobe filling supergiants with short spin periods and underfilled Roche lobe supergiants 
with longer spin periods \citep{jenke2012}. A majority of supergiant HMXBs are persistent sources. Some of the supergiant HMXBs, 
called the highly absorbed supergiant HMXBs, have very strong absorption with column density of absorbing matter 
N$_{H} \sim 10^{23}$ cm$^{-2}$.
\par
IGR J16393-4643 belongs to this increasing class of heavily absorbed HMXBs detected by the {\it INTEGRAL} survey of the galactic plane, 
that are mainly concentrated in the spiral arms. It was first discovered with the {\it ASCA} observatory during survey of 
the galactic plane \citep{sugizaki2001} and was named as AX J1639.0-4642. It has a spin period of $\sim$ 910 s and a 4.2 days orbital period \citep{bodaghee2006,corbet2010} 
and occupies a unique position near the top edge of underfilled Roche lobe supergiant systems in the Corbet diagram \citep{jenke2012}. 
Due to the lack of detailed studies of such systems, IGR J16393-4643 makes an interesting candidate to study the timing and spectral properties of the 
short orbital period and long spin period supergiant systems. 
\par
The X-ray spectrum of the pulsar is characterised by a highly absorbed power-law with an exponential 
cut-off along with Fe fluorescence lines \citep{lutovinov2005,bodaghee2006}. The strong absorption of the order $\sim 10^{23}$ cm$^{-2}$ is an evidence of a dense circumstellar 
environment surrounding the pulsar. Previous observations with {\it XMM-Newton} indicates the presence 
of a soft excess in the spectrum which could be due to X-rays scattering by the stellar wind \citep{bodaghee2006}.
\par
Here we present the simultaneous broad-band pulse profiles and spectral characteristics of the pulsar IGR J16393-4643 obtained from a {\it Suzaku} observation. We also 
present results from the orbital intensity profile analysis of IGR J16393-4643 using {\it SWIFT-BAT} light-curves. 
The energy resolved pulse profiles are created for the first time for this source. 
The broad-band spectral characteristics are studied both for the phase averaged as well as resolving the pulse phase into peak phase and trough phase. 
These results provide valuable insights into the nature of the such underfilled Roche lobe supergiant systems.

\section{Observations and Data Analysis}
{\it Suzaku} is the fifth Japanese X-ray astronomy satellite launched in July 2005. It consists of two sets of co-aligned scientific 
instruments, the X-ray Imaging Spectrometer (XIS), operating in the energy range 0.2-12 keV and Hard X-ray Detector (HXD), operating 
in the energy range 10-600 keV. The XIS consists of three front illuminated CCD detectors and one back illuminated CCD detector 
\citep{koyama2007}, out of which three CCDs XIS0, XIS1 and XIS3 are currently operational. The HXD consists of silicon PIN diodes 
operating in energy range 10-70 keV and GSO crystal scintillators extending the energy range till 600 keV \citep{takahashi2007}.
\par
IGR J16393-4643 was observed with {\it Suzaku} during 12 March 2010 (ObsId: 404056010) with an useful exposure time of $\sim$ 44 kilosecs over a span of about 120 kilosecs. 
The observations were carried out at the `XIS nominal' pointing position and the XIS were operated in `standard' data mode in the 
`Normal window' option, having a time resolution of 8 secs. For both the XIS and HXD data, we have used the filtered cleaned event files 
which are obtained using the pre-determined screening criteria described in the Suzaku ABC guide.
\footnote{http://heasarc.nasa.gov/docs/suzaku/analysis/abc/node9.html}
\par
The XIS light-curves and spectra were extracted from the cleaned event files by selecting circular regions of 3' around the 
source centroid. The background light-curves and spectra were extracted by selecting circular regions of same size away from the 
source centroid. For HXD/PIN background, simulated `tuned' non X-ray background event files (NXB) corresponding to the month and year of 
the observation was used to estimate the non X-ray background.\footnote{http://heasarc.gsfc.nasa.gov/docs/suzaku/analysis/pinbgd.html} 
The cosmic X-ray background was simulated as suggested by the instrument team with 
appropriate normalisations and response files.\footnote{http://heasarc.gsfc.nasa.gov/docs/suzaku/analysis/pin\_cxb.html} 
The response files for XIS were created using CALDB `20140211' and for HXD/PIN, response files were obtained from the 
\emph{Suzaku} Guest Observer Facility.\footnote{http://heasarc.gsfc.nasa.gov/docs/heasarc/caldb/suzaku} 

\section{Timing Analysis}

\begin{figure*}
\centering
\includegraphics[angle=-90,scale=0.4]{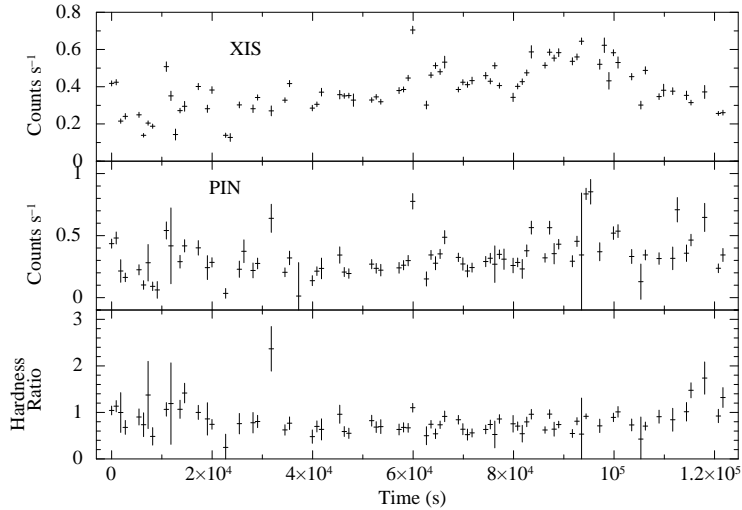} 
\caption{Light-curve of IGR J16393-4643 binned at the pulse period of 908 secs in 0.3-12 keV XIS energy band (top panel), 12-50 keV 
PIN energy band (middle panel) and the hardness ratio of the count-rates in PIN and XIS (bottom panel) are shown here.}
\label{lightcurves}
\end{figure*}

For the timing analysis of the source, we have applied barycentric corrections to the event data files using the {\small FTOOLS} 
task `aebarycen'. Light-curves with time resolution of 1 s and 8 s were extracted from HXD/PIN (12-50 keV) and XIS (0.3-12 keV) 
respectively. Figure.~\ref{lightcurves} shows the light-curves binned with a time bin of 908 secs {\it i.e} at the pulsar spin period, 
in XIS and PIN along with the hardness ratio. The count-rate in XIS and PIN increases gradually from the start of the observation upto 
100 kilosecs by a factor of $\sim$ 2 and then decreases by a similar factor till the end of the observation (also see second and third panel in Figure~\ref{orbital}). 
The hardness ratio remained constant throughout the observation. 

\subsection{Orbital intensity profile analysis}

\begin{figure*}
\centering
\includegraphics[angle=-90,scale=0.45]{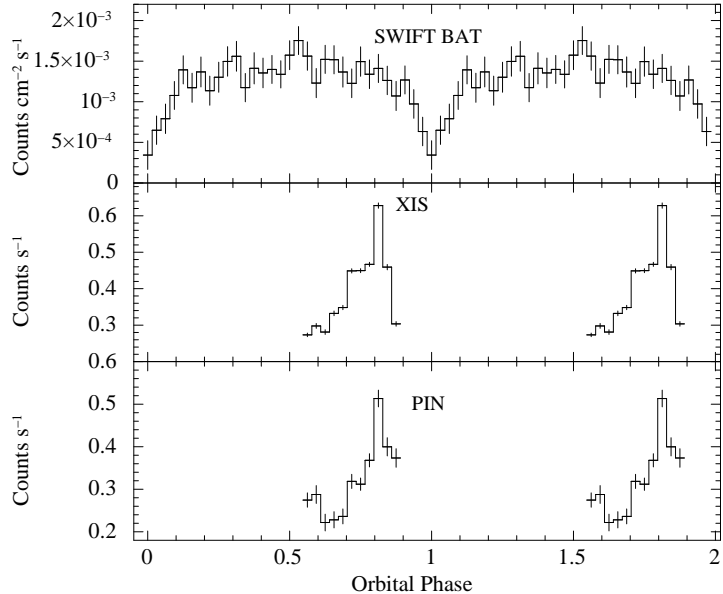} 
\caption{Light-curve of {\it SWIFT BAT}, XIS and PIN folded with the orbital period (P$_{orb}$ = 366150 secs) of IGR J16393-4643. The orbital phase zero corresponds to the eclipse at 
epoch MJD: 53417.955}
\label{orbital}
\end{figure*}

We have used long term light-curve of IGR J16393-4643 in 15-50 keV {\it SWIFT-BAT} energy band to estimate the orbital period of the system 
(P$_{orb}$ = 366150 secs; consistent with \citealt{corbet2010}). We then folded the XIS and PIN light-curves with the {\it SWIFT-BAT} 
light-curves, to investigate the orbital phase of the {\it Suzaku} observation. The minimum of the orbital intensity profile is taken as orbital phase zero, corresponding to 
epoch MJD: 53417.955. Figure~\ref{orbital} shows an indication that the system is an eclipsing binary, which was also reported previously by \cite{corbet2013}. 
The eclipse duration is short with the eclipse semi-angle $\theta_{E} \sim$ 17$^{\circ}$. 
The {\it Suzaku} observations were carried out roughly from orbital phase 0.55-0.9 {\it i.e} prior to going into the eclipse.
As seen in Figure~\ref{orbital}, the count-rate in XIS and PIN increases by a factor of $\sim$ 2 during orbital phase $\sim$ 0.65-0.8 and then again decreases till the 
end of the observation. The orbital intensity profile created with {\it SWIFT-BAT} is averaged over many orbital cycles, whereas the orbital intensity profile created 
with XIS and PIN show variability on sub-orbital timescales, similar to that seen in OAO 1657-415 \citep{pradhan2014}.

\subsection{Energy resolved pulse profiles}

\begin{figure*}
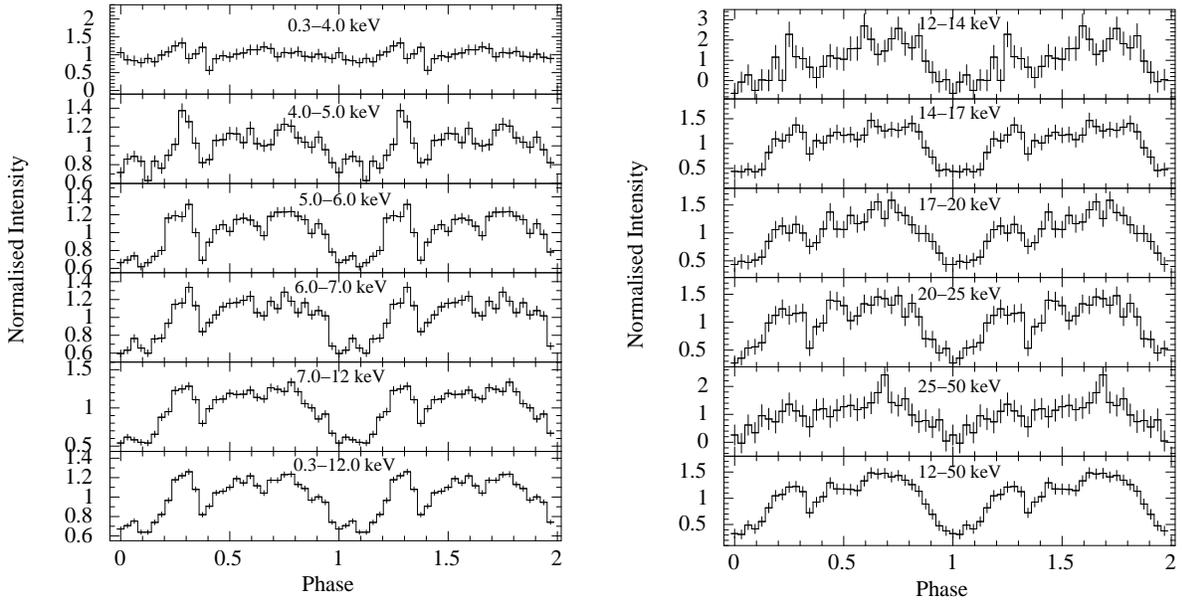

\centering
\includegraphics[angle=-90,scale=0.45]{xis_profiles_replot.ps} 
\includegraphics[angle=-90,scale=0.45]{pin_profiles_replot.ps}
\caption{Energy resolved pulse profiles of IGR J16393-4643 folded with a period of 908.79 secs for XIS energy bands (left panel) and PIN energy bands (right panel). 
The epoch is chosen such that the pulse minima occur at phase 0.0 in the PIN 12-50 keV energy band.}
\label{pulseprofiles}
\end{figure*}

\begin{figure*}
\centering
\includegraphics[angle=-90,scale=0.45]{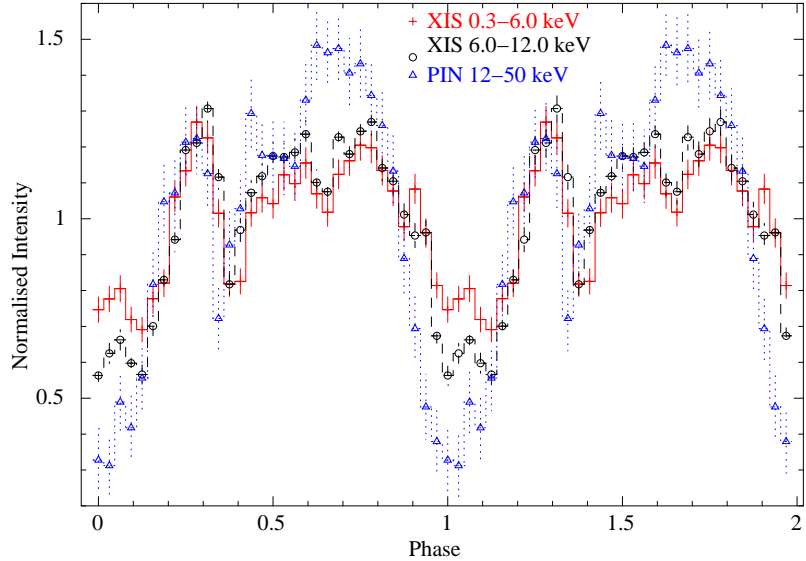} 
\caption{Overlaid pulse profiles created in XIS energy band 0.3-6.0 keV and 6-12 keV and in PIN energy band 12-50 keV, shows an indication of a phase lag.}
\label{xis_pin}
\end{figure*}

We have searched for pulsations in the light-curves by applying pulse folding and $\chi^{2}$ maximization technique and the pulse period 
was found to be 908.79 $\pm$ 0.01 secs. We then created the energy resolved pulse profiles by folding light-curves extracted in different energy 
bands with the pulse period. Light-curves from XIS0, XIS1 and XIS3 were added together to create the pulse profiles in XIS 
energy band 0.3-12 keV and sub bands within. From Figure~\ref{pulseprofiles}, we see that the pulse profiles 
have a complex morphology with some energy dependance, which was also seen with {\it XMM-Newton} \citep{bodaghee2006} and 
{\it RXTE PCA} \citep{thompson2006}, but the broad-band nature of the pulse profiles are brought out from this {\it Suzaku} observation. 
Due to the presence of high column density of absorbing matter, very few photons are detected below 4 keV. 
The epoch for creation of the pulse profiles is adjusted to make the minima appear at phase zero in the PIN 12-50 keV energy band. 
The pulse minima is seen to become deeper at higher energies. A phase lag is seen in the energy resolved pulse profiles in the two XIS energy bands compared to the PIN energy bands, 
which is shown in Figure~\ref{xis_pin} for pulse profiles created in XIS energy band 0.3-6.0 keV and 6-12 and in PIN energy band 12-50 keV.

\section{Spectral Analysis}

\subsection{Pulse phase averaged spectrum}
We have performed the pulse phase averaged spectral analysis of IGR J16393-4643 using spectra from three XIS and PIN. The energy range chosen for spectral fitting was 
3.0 keV to 10.0 keV for XIS, since there were hardly any photons below 3.0 keV. For PIN, the energy range was chosen from 15 keV to 50 keV. The XIS spectra were rebinned by a 
factor of 12 upto 5 keV, by a factor of 6 from 5-7 keV and again by a factor of 12 from 7-10 keV. The PIN spectra was rebinned by a factor of 2 till 22 keV, by 8 from 22-45 keV 
and by a factor of 12 for the rest. The spectral fitting was done with {\small XSPEC} 12.8.1. To fit the continuum spectra, we have tried several standard continuum models used for 
HMXB pulsars like cut-off power-law (CUTOFFPL), power-law with high energy cut-off (HIGHECUT, \citealt{white1983}), power-law with a Fermi-Dirac cut-off 
(FDCUT, \citealt{tanaka1986}), and negative and positive power-law exponential (NPEX, \citealt{mihara1995}). 
The power-law with HIGHECUT and power-law with FDCUT spectral models provided the best fits to the phase averaged spectrum with physically acceptable parameter values and 
henceforth, we report results from these two spectral models.
\par
\begin{figure*}
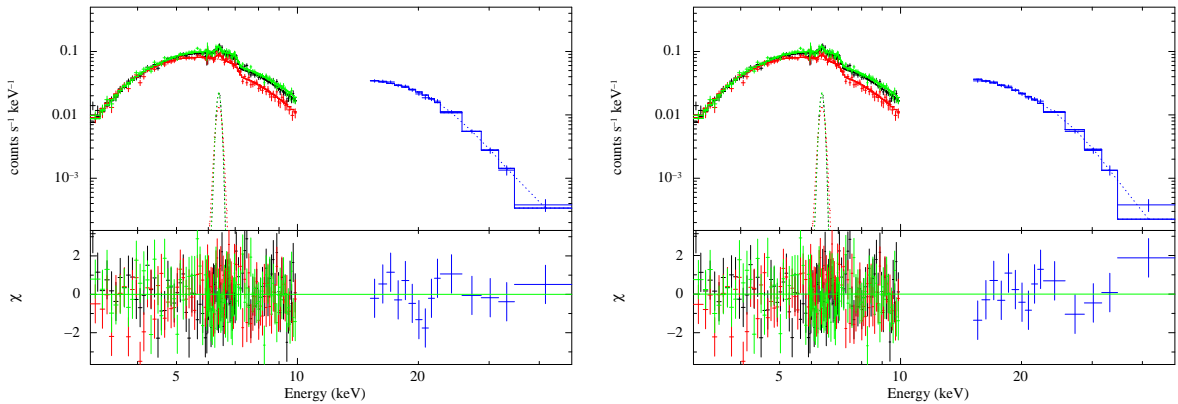

\centering
\includegraphics[angle=-90,scale=0.3]{highecut_spectra.ps}
\includegraphics[angle=-90,scale=0.3]{fdcut_spectra.ps}
\caption{Phase averaged spectrum of IGR J16393-4643 with the best fit models power-law with HIGHECUT (left panel) and power-law with FDCUT (right panel) are shown here 
along with the contribution of residuals to the $\chi^{2}$.}
\label{spectra}
\end{figure*}

We fitted the spectra simultaneously with all the parameters tied, except the relative normalisations of the instruments which were kept free. 
A 6.4 keV Fe K$\alpha$ line was also found in the spectra, which was modelled by a gaussian emission line. The 7.1 keV K$\beta$ line detected in the spectra with the 
{\it XMM-Newton} observation \citep{bodaghee2006} was marginally detected here with the 90 $\%$ upper limit on the line flux as 
$2 \times 10^{-5}$ photons cm$^{-2}$ s$^{-1}$ and equivalent width of 33 eV, consistent with the upper limits quoted for {\it XMM-Newton} 
observation ($6 \times 10^{-5}$ photons cm$^{-2}$ s$^{-1}$ and 120 eV respectively; \citealt{bodaghee2006}). 
The best fit parameter values of phase averaged spectrum for 90$\%$ confidence limits (except for $\sigma_{K\alpha}$ and equivalent width of the Fe line 
which are given for 1$\sigma$ confidence limits) for the two spectral models power-law with HIGHECUT and power-law with FDCUT are given in Table 1. Figure~\ref{spectra} shows the 
best fit phase averaged spectrum using power-law with HIGHECUT and power-law with FDCUT models, along with the residuals.

\begin{table*}
\label{table}
 \caption{Best fit parameter values for power-law with HIGHECUT and power-law with FDCUT models for the phase averaged spectrum as well as the peak phase and trough phase spectra. 
Errors quoted for $\sigma_{K \alpha}$ and equivalent width of the gaussian Fe K$\alpha$ line are for 1 $\sigma$ confidence limits and for rest of the parameters are for 90$\%$ confidence limits}
\centering
\begin{tabular}{|c c c c c c c|}
\hline
Parameters & Phase averaged & & Peak phase & & Trough phase & \\
\hline 
 & HIGHECUT & FDCUT & HIGHECUT & FDCUT & HIGHECUT & FDCUT \\
\hline
\vspace{3mm}
N$_{H}$ ($10^{22}$ atoms cm$^{-2}$) & $26.5 \pm 0.8$ & $26.2 \pm 0.8$ & $27.4 \pm 1.2$ & $27.1 \pm 1.2$ & $26.4 \pm 1.7$ & $26.0^{+2}_{-3}$ \\
\vspace{3mm}
Photon Index ($\Gamma$) & $0.91 \pm 0.06$ & $0.86^{+0.07}_{-0.08}$ & $0.82 \pm 0.09$ & $0.76^{+0.10}_{-0.13}$ & $1.2 \pm 0.1$ & $1.2 \pm 0.1$ \\
\vspace{3mm}
$\Gamma_{norm}$ ($10^{-3}$ photons cm$^{-2}$ s$^{-1}$) & $3.8^{+0.5}_{-0.4}$ & $3.5 \pm 0.5$ & $3.7^{+0.8}_{-0.6}$ & $3.4^{+0.8}_{-0.6}$ & $5.1^{+1.6}_{-1.2}$ & $4.7^{+1.8}_{-1.4}$ \\
\vspace{3mm}
Cut-off Energy (E$_{C}$ keV) & $20 \pm 1$ & $25^{+1}_{-3}$ & $20 \pm 1$ & $25^{+2}_{-5}$ & $19^{+6}_{-4}$ & $25^{+4}_{-6}$ \\
\vspace{3mm}
Fold Energy (E$_{F}$ keV) & $9 \pm 1$ & $5 \pm 1$ & $9 \pm 2$ & $5 \pm 2$ & $11^{+5}_{-7}$ & $6 \pm 3$ \\
\vspace{3mm}
F$_{K \alpha}$ ($10^{-5}$ photons cm$^{-2}$ s$^{-1}$) & $3.2^{+0.8}_{-0.7}$ & $3.2^{+0.8}_{-0.7}$ & $3.9 \pm 1.1$ & $3.9 \pm 1.1$ & $5.1 \pm 2.0$ & $5.2^{+1.9}_{-2.1}$ \\
\vspace{3mm}
$\sigma_{K \alpha}$ (keV) & $0.02^{+0.03}_{-0.02}$ & $0.02^{+0.03}_{-0.02}$ & - & - & $0.13^{+0.05}_{-0.06}$ & $0.14^{+0.04}_{-0.06}$ \\
\vspace{3mm}
Equivalent width (eV) & $46^{+7}_{-6}$ & $46^{+7}_{-6}$ & $47^{+8}_{-8}$ & $48^{+8}_{-8}$ & $99^{+23}_{-22}$ & $102^{+21}_{-25}$ \\
\vspace{3mm}
Flux (XIS) (0.3-12 keV) ($10^{-11}$ ergs cm$^{-2}$ s$^{-1}$) & $3.6 \pm 0.1$ & $3.5 \pm 0.1$ & $4.2 \pm 0.1$ & $4.2 \pm 0.1$ & $2.5 \pm 0.1$ & $2.5 \pm 0.1$ \\
\vspace{3mm}
Flux (PIN) (12-70 keV) ($10^{-11}$ ergs cm$^{-2}$ s$^{-1}$) & $2.7 \pm 0.1$ & $2.6 \pm 0.1$ & $3.4 \pm 0.2$ & $3.2 \pm 0.2$ & $1.6 \pm 0.1$ & $1.5 \pm 0.2$ \\
\vspace{3mm}
$\chi^{2}_{\nu}$/d.o.f & 1.17/296 & 1.18/296 & 1.03/280 & 1.05/281 & 1.10/289 & 1.09/289 \\
\hline
\end{tabular}
\end{table*}

\subsection{Pulse phase resolved spectrum}
\begin{figure*}
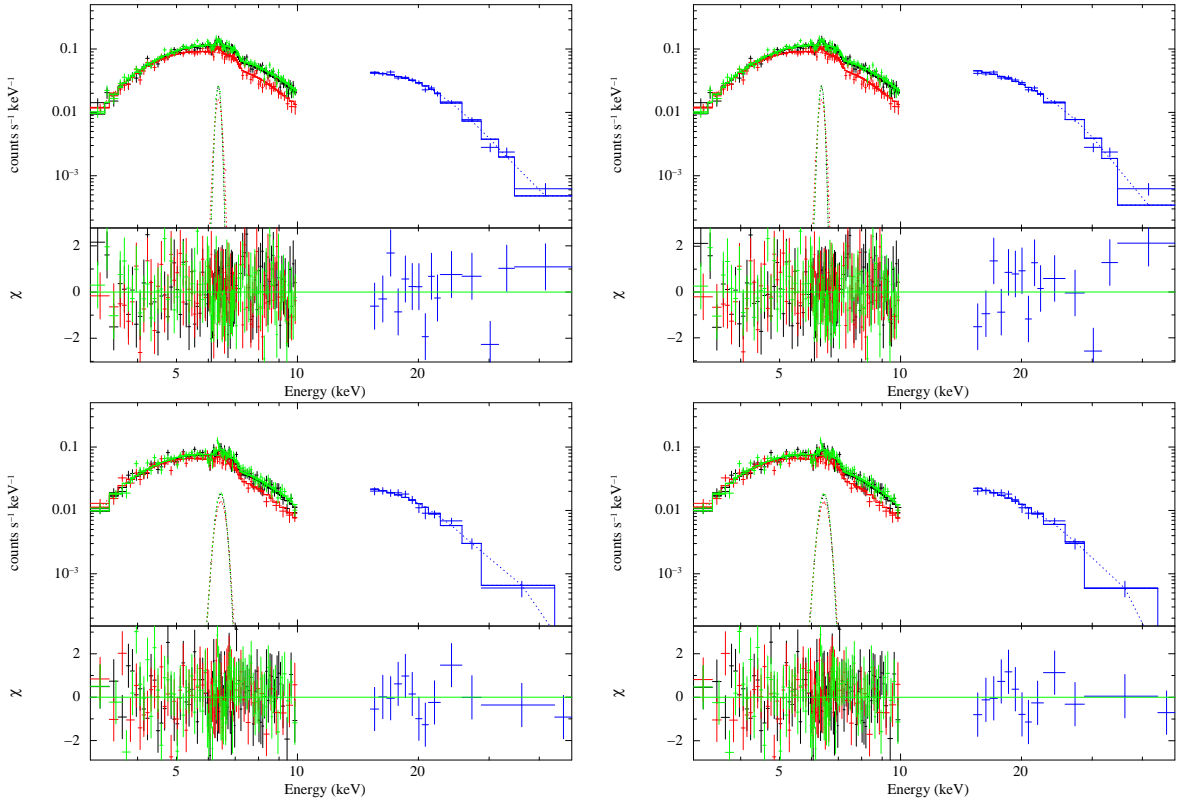

\centering
\includegraphics[angle=-90,scale=0.3]{highecut_pulse.ps} 
\includegraphics[angle=-90,scale=0.3]{fdcut_pulse.ps} \\
\includegraphics[angle=-90,scale=0.3]{highecut_plateau.ps} 
\includegraphics[angle=-90,scale=0.3]{fdcut_plateau.ps}
\caption{Peak phase spectrum of IGR J16393-4643 using power-law with HIGHECUT model (upper left panel) and power-law with FDCUT model (upper right panel) are shown 
here, along with the contribution of residuals to the $\chi^{2}$. Same for the trough phase are shown in the two bottom panels.}
\label{phaseresolved}
\end{figure*}

The energy dependence of the pulse fraction as seen in the XIS and PIN energy bands (Figure~\ref{pulseprofiles} and Figure~\ref{xis_pin}) indicates a significant change in the 
spectrum with the pulse phase and we investigate the same with pulse phase resolved spectroscopy at the peak and trough of the pulse profile. The XIS and PIN spectrum were 
binned into two states: a peak phase around the pulse maximum (phase 0.4-0.8) and a trough phase around the pulse minima (phase 0.0-0.2 and 0.9-1.0), 
similar to the pulse phase definition used in \citet{bodaghee2006}. 
In the peak phase as well as the trough phase, the XIS spectra and PIN spectra were rebinned by the same factors used in phase averaged spectra. 
To study the changes in spectral parameters in the two pulse phases, we used the same models as in the phase averaged spectrum. 
However, width of the Fe K$\alpha$ line could not be constrained in the peak phase spectra and was fixed to its phase averaged value. 
The best fit parameter values of the peak phase spectra and the trough phase spectra for 90$\%$ confidence limits (except for $\sigma_{K\alpha}$ and equivalent width of the 
Fe line for both the spectral models which are given for 1$\sigma$ confidence limits) for the two spectral models power-law with HIGHECUT and power-law with FDCUT are given in Table.1. 
Figure~\ref{phaseresolved} shows the peak phase and trough phase spectra using power-law with HIGHECUT and power-law with FDCUT spectral models, along with the residuals. 
A softening in the spectrum with an increase in the Fe equivalent width is noticed at the trough phase.

\section{Discussions}

\subsection{Orbital intensity profile and inclination of the system}

The first panel in Figure~\ref{orbital} is the orbital intensity profile created by folding the long term light-curve of IGR J16393-4643 obtained with {\it SWIFT-BAT} in 
energy range 15-50 keV, with the orbital period of the system (P$_{orb}$ = 366150 secs). The orbital intensity profile shows a narrow eclipse around orbital phase zero with 
the eclipse semi-angle $\theta_{E} \sim$ 17$^{\circ}$. Recent {\it Chandra} observation of this binary system proposed the optical counterpart to be either 
a supergiant O9 star with radius R$\sim$ 20 R$_{\odot}$ or a main sequence B star with radius R$\sim$ 10 R$_{\odot}$ \citep{bodaghee2012}. 
The short orbital period of the binary system makes it more likely to have a supergiant O9 star as its companion than a main sequence B star. 
However, \cite{bodaghee2012} pointed out that having a supergiant O9 star as a companion would imply a lower limit of distance of 25 kpc to the binary as compared to a distance 
of 12 kpc for a main sequence B star.
\par
The supergiant HMXBs like Cen X-3, Vela X-1 and OAO 1657-415 have masses of the companion star in the range of 10 M$_{\odot}$ - 30 M$_{\odot}$ 
and radii in the range of 10 R$_{\odot}$ - 35 R$_{\odot}$, with a nearly circularised orbit \citep{rappaport1983,ash1999,van1995,mason2012}. 
So assuming a circular orbit, the separation between the binary components {\it a}, can be expressed as 
\begin{equation}
\label{kepler}
 a^{3}=\frac{G(M_{c}+M_{NS})P_{orb}^{2}}{4\pi^{2}}
\end{equation}
where $\mathrm {M_{NS}}$ is the mass of the neutron star which is assumed to be 1.4 M$_{\odot}$, $\mathrm {P_{orb}}$ = 4.24 days is the orbital period of the system and 
$\mathrm {M_{c}}$ is the mass of the supergiant companion star which is assumed to be in the range 10 M$_{\odot}$ - 30 M$_{\odot}$. \\
For an orbital separation {\it a}, we have then calculated the range of inclination $\mathrm {i}$ as a function of the companion star radius $\mathrm{R_{O}}$, 
given by Equation~\ref{equation} \citep{rubin1996}.
\begin{equation}
\label{equation}
 \mathrm{\frac{R_{O}}{a} = \sqrt{cos^{2}i+sin^{2}i \quad sin^{2}\theta_{E}}}
\end{equation}
where $\theta_{E}\sim$ 17$^{\circ}$ is the observed eclipse semi-angle. The upper limit to the companion radius for a given mass is constrained by the Roche lobe radius 
R$_{Roche}$, which is given by the Equation~\ref{roche} \citep{bowers1984,hill2005}
\begin{equation}
 \label{roche}
 R_\mathrm{{Roche}} = a (0.38+0.2 \mathrm{log}(\frac{M_{c}}{M_{NS}}))
\end{equation}

\begin{figure*}
\centering
\includegraphics[angle=-90,scale=0.3]{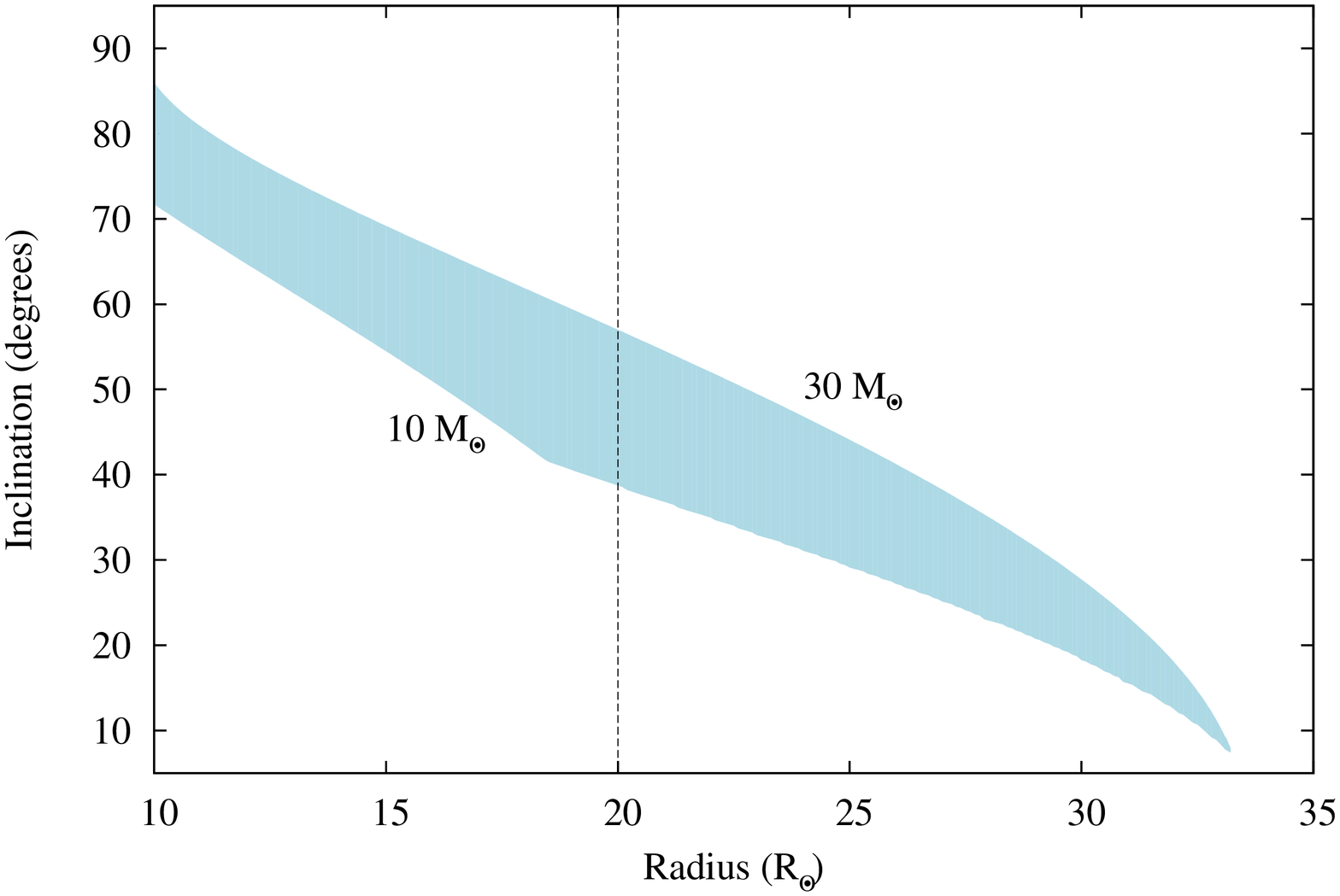} 
\includegraphics[angle=-90,scale=0.3]{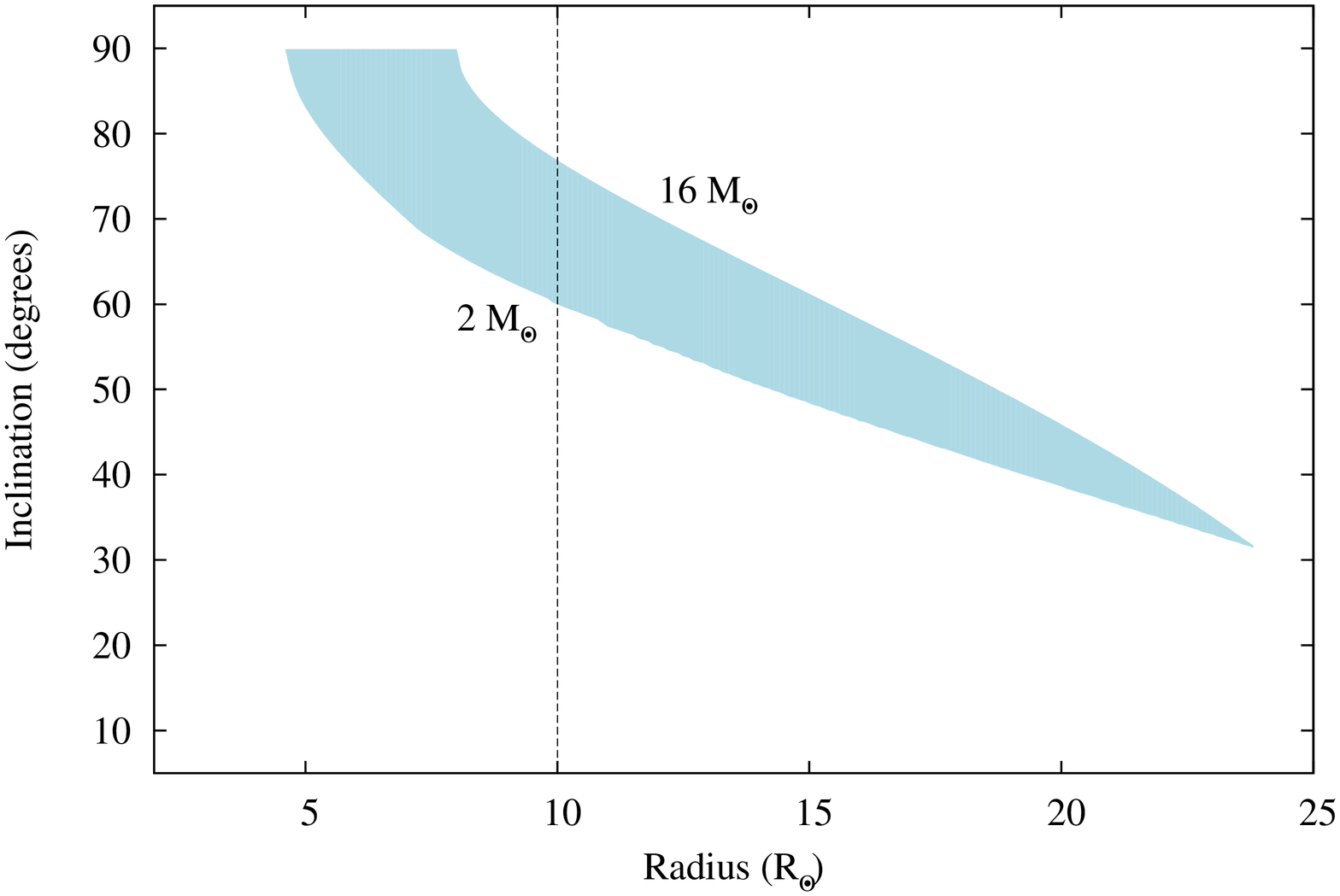} 
\caption{Left panel shows the plot of the range of inclination angles allowed for given mass of the O9 supergiant companion star as a function of its radius, assuming a circular orbit. 
Right panel shows the plot of the range of inclination angles allowed for given mass of the main sequence B star as a function of its radius, also assuming a circular orbit. 
See text for details. The dotted lines represent the 20 R$_{\odot}$ radius for a supergiant companion and the 10 R$_{\odot}$ radius for a B type companion (Bodaghee et al. 2012).}
\label{inclination}
\end{figure*}

\par
Similiarly, we carried out the above calculations for the companion as a main sequence B star having masses in the range of 2 M$_{\odot}$ - 16 M$_{\odot}$ 
and radii in the range of 5 R$_{\odot}$ - 25 R$_{\odot}$ \citep{habets1981} and assuming a circular orbit. Figure~\ref{inclination} is the plot of the range of inclination angles 
allowed for given mass of the O9 supergiant companion star and main sequence B star as a function of its radius, assuming a circular orbit. 
\par
For a supergiant companion star radius of 20 R$_{\odot}$, the inclination of the orbital plane of 
the binary system is in the range $39^{\circ}-57^{\circ}$. For a main sequence B star of radius 10 R$_{\odot}$, the inclination of the orbital 
plane of the binary system is in the range $60^{\circ}-77^{\circ}$. 

\subsection{Energy dependent pulse profiles}

Figure~\ref{pulseprofiles} shows the complex double peaked pulse profiles of IGR J16393-4643 in the {\it Suzaku} energy sub-bands. The pulse fraction increases with energy, 
from $\sim$ 33$\%$ in XIS 0.3 - 12 keV energy band to $\sim$ 65$\%$ in PIN 12 - 50 keV energy band, which is also seen in many other X-ray pulsars \citep{white1983,frontera1989,nagase1989}. 
There is a narrow dip present in the pulse profiles at pulse phase $\sim$ 0.35, which was also seen in pulse profiles created with {\it RXTE PCA} \citep{thompson2006} and 
{\it XMM-Newton} \citep{bodaghee2006}. This narrow dip exists even at higher energies as seen in Figure~\ref{pulseprofiles} and Figure~\ref{xis_pin} for XIS and PIN energy bands. 
Narrow high energy dips in the pulse profiles are seen in other X-ray pulsars like GX 1+4 \citep{naik2005}, EXO 2030+375 \citep{naik2013}, A 0535+262 \citep{frontera1985,cemeljic1998} 
and are associated with the intrinsic beaming pattern of the source. 
From Figure.~\ref{xis_pin}, we see a phase lag of the soft photons with respect to the hard photons in the energy resolved pulse profiles. 
Such soft phase lags have been observed in millisecond X-ray pulsars \citep{cui1998,ibragimov2011}. 
However, further analysis with respect to the actual cause of phase lag is limited by the statistical quality of the data.

\subsection{Phase resolved spectral characteristics}

From phase averaged spectral characteristics of IGR J16393-4643 (Table 1 and Figure~\ref{spectra}), we find very high line of sight column density of absorbing matter 
$\sim 3 \times 10^{23}$ cm$^{-2}$. Such high column density of absorbing matter has been found in previous studies of this system by {\it RXTE-PCA} \citep{thompson2006} 
and by {\it XMM-Newton} \citep{bodaghee2006} and is attributed to the circumstellar environment around the pulsar. 
The phase averaged spectrum is described by a power-law with spectral index $\Gamma \sim$ 0.9 and a high energy cut-off above 20 keV. 
The presence of the soft excess as detected from the {\it XMM-Newton} observation \citep{bodaghee2006} could not be confirmed in the {\it Suzaku} observation. 
Due to limited statistical quality of this {\it Suzaku} observation, it is difficult to make an in-depth pulse phase resolved spectral analysis. 
\par
Instead, the pulse profile is broadly resolved into peak phase and trough phase and pulse phase dependance of the spectral properties are studied in these two 
phase-bins (Figure~\ref{phaseresolved}). While the column density is similar in both the pulse phases, there is a change in the continuum spectral parameters which is observed for 
both the spectral models in Table 1. 
The spectrum is softer at the trough phase and harder at the peak phase, which may be due the additional softer photons near the off pulse regions.
Alternately, this may also imply a deeper and more direct view into the emission region along the magnetic axis at the pulse peak as would be for the case for a fan beam kind of 
emission pattern \citep{pravdo1976}. The cut-off parameters (fold energy E$_{F}$ and cut-off energy E$_{C}$) for both power-law with HIGHECUT and power-law with 
FDCUT spectral models however, remain constant in the two pulse phases. The Fe K$\alpha$ line is present in both the phases, with the equivalent width higher in 
trough phase than in peak phase. In contrast with the results obtained in \cite{bodaghee2006} with the {\it XMM-Newton}, which operated in the limited energy band of 
0.3-10 keV, there is an underlying change in the source spectrum as a function of pulse phase which is brought out in this broad-band {\it Suzaku} observation. 
\par
IGR J16393-4643 makes an interesting candidate for detailed pulse phase resolved spectroscopy with future X-ray missions with broad-band energy coverage. Such in-depth analysis 
will help in better understanding of the accretion geometry and beaming pattern of these underfilled Roche lobe systems.
\par
\vspace{15mm}
\textbf{ACKNOWLEDGEMENT}\\
The authors thank an anonymous referee for useful comments. The data used for this work has been obtained through the High Energy Astrophysics Science Archive (HEASARC) 
Online Service provided by NASA/GSFC. We have also made use of the public light-curves from {\it SWIFT-BAT} site.

\end{document}